\documentclass[structabstract]{aa}
\usepackage{graphicx}
\usepackage{txfonts}
\begin{document}
   \title{Fundamental stellar and accretion disc parameters of the eclipsing binary DQ Velorum}


   \author{D. Barr\'ia \inst{1,2}, R. E. Mennickent \inst{1}, L. Schmidtobreick \inst{2}, G. Djura\v{s}evi\'{c} \inst{3,4},
  Z. Ko\l{}aczkowski \inst{5},  G. Michalska \inst{5}, M. Vu\v{c}kovi\'{c} \inst{2,3}, E. Niemczura \inst{6}}

   \institute{Universidad de Concepci\'on, Departamento de Astronom\'ia, Concepci\'on, Chile\\
              \email{dbarria@astro-udec.cl,rmennick@astro-udec.cl}
         \and
             European Southern Observatory, Vitacura, Santiago, Chile\\
             \email{lschmidt@eso.org, mvuckovi@eso.org}
         \and
             Astronomical Observatory, Belgrade, Serbia\\
             \email{gdjurasevic@aob.rs}
         \and
             Isaac Newton Institute of Chile, Yugoslavia Branch\\
        \and
             Instytut Astronomiczny Uniwersytetu Wroclawskiego, Wroc\l{}aw, Poland\\
             \email{zibi@astro-udec.cl, michalska.gabi@gmail.com}
         \and
             Astronomical Institute, Wroc\l{}aw University, Wroc\l{}aw, Poland \\
             \email{eniem@astro.uni.wroc.pl}}


 
  \abstract
   {To add to the growing collection of well-studied double periodic variables (DPVs) we have carried out
   the first spectroscopic and photometric analysis of the eclipsing binary DQ Velorum to obtain its main physical stellar and orbital parameters.}
   {Combining spectroscopic and photometric observations that cover several orbital cycles allows us to estimate the stellar properties of the
    binary components and the orbital parameters. We also searched for circumstellar material around the more massive star.}
   {We separated DQ Velorum composite spectra and measured radial velocities with an iterative method for double spectroscopic binaries. We obtained the radial
velocity curves and calculated the spectroscopic mass ratio. We compared our single-lined spectra with a grid of synthetic spectra
and estimated the temperature of the stars. We modeled the V-band light curve with a fitting method based on the simplex algorithm, which includes an
accretion disc.
 To constrain the main stellar parameters we fixed the mass ratio and donor temperature to the values obtained by our spectroscopic analysis.}
   {We obtain a spectroscopic mass ratio $q=0.31\pm0.03$ together with donor and gainer masses $M_\mathrm{d}=2.2\pm0.2\ M_{\odot}$, $M_\mathrm{g}=7.3\pm0.3\ M_{\odot}$, the radii 
 $R_\mathrm{d}=8.4\pm0.2\ R_{\odot}$, $R_\mathrm{g}=3.6\pm0.2\ R_{\odot}$ and temperatures $T_\mathrm{d}=9400\pm100\ \mathrm{K}$, $T_{g}=18\,500\pm500\  \mathrm{K}$ for
 the stellar components. 
 We find that DQ Vel is a semi-detached system consisting of a B3V gainer and an A1III donor star plus an extended accretion disc around the gainer. 
 The disc is filling $89\%$ of the gainer Roche lobe with a temperature of $6580\pm300\ \mathrm {K}$ at the outer radius. It has a concave shape that is thicker 
 at its edge ($d_\mathrm{e}=0.6\pm0.1\ R_{\odot}$) than at its centre ($d_\mathrm{c}=0.3\pm0.1\ R_{\odot}$). 
 We find a significant sub-orbital frequency of $0.19\,d^{-1}$ in the residuals of the V-band light curve, which we interpret as a pulsation
 of an slowly pulsating B-type (SPB) of a gainer star.
 We also estimate the distance to the binary ($d\sim3.1\ \mathrm{kpc}$) using the absolute radii, apparent magnitudes, and effective temperatures of the components found 
in our study.}
   {}

  \titlerunning{Fundamental stellar and accretion disc parameters of the eclipsing binary DQ Velorum}
  \authorrunning {Barr\'ia et al.}
   \maketitle


\section{Introduction}

 Discovered by Hoffmeister (\cite{hoffmeister}), DQ Velorum (TYC 8175-333-1, ASAS J093034-5011.9) is a Galactic eclipsing binary of Algol-type with an orbital
 period of 6.08337 d (Hoffmeister \cite{hoffmeister}). Considering that DQ Vel is a bright system (V$\sim10.7$), it is still poorly analysed
 in detail.\\
 A study of the system using photographic plates was made by van Houten (\cite{vanhouten}), who determined the following ephemeris for the primary minimum:
 \begin{equation}
  T_\mathrm{min} = \mathrm{JD}\,2430881.20(3) + 6.08337(13)\times E\,.
 \end{equation}
 Milone (\cite {milone}), using UVB photometry, indicated that DQ Vel shows a $\beta$ Lyrae-like light curve and investigated the O'Connell effect on the system 
 (O'Connell \cite{oconnell}), studying the differential light between maxima of the light curves. He concluded that the O'Connell effect is positive and
 stronger in the U band.
 He also used low-resolution spectra obtained at the Cerro Tololo inter-american observatory with the image-tube spectrograph at the Yale 1-m telescope to obtain an estimate of the
 spectral type as A3V. However, no attempt was made to measure the radial velocities.\\
 More recently, Michalska et al. (\cite{michalska}) discovered an additional long-term variability of 188.9 d using ASAS photometric data.
 This places the system in the new group of variable stars called \emph{double periodic variables} (hereafter DPVs). 
 This group of interacting binaries is characterised by two photometric variabilities
 linked one to each other with a period ratio of about 33 (Mennickent et al. \cite{mennickent2003}, Mennickent \& Ko\l{}aczkowski \cite{mennickent2010a}). 
 The short-term variability corresponds to the orbital motion of the binary 
 while the long-term variability is still not completely understood.
 DPVs have been interpreted as semi-detached binaries showing cycles of mass loss into the 
 interstellar medium (Mennickent et al. \cite{mennickent2008}, \cite{mennickent2012b}).\\
 To study these interesting objects, an intense observational effort is being carried out 
 to find DPVs both in the Galaxy and the Magellanic Clouds using the microlensing surveys OGLE and MACHO (Mennickent et al. \cite{mennickent2005}, 
 Poleski et al. \cite{poleski}), as well as in the ASAS project.
 At present, there are more than 150 DPVs discovered in the Magellanic Clouds and 11 systems (including DQ Vel) in the Milky Way 
 (Mennickent et al. \cite{mennickent2012a}).
 Results of multiwavelength spectroscopy and/or photometry have made it possible to study DPVs in more detail,
 such as AU Mon (Desmet et al. \cite{desmet}, Djura\v{s}evi\'{c} et al. \cite{djurasevic}), 
 V393 Sco (Mennickent et al. \cite{mennickent2010b}, \cite{mennickent2011a}), and LP Ara (Mennickent et al. \cite{mennickent2011b}).\\
 Here, we present the first optical spectroscopic analysis of DQ Vel using 46 high-resolution spectra that cover several orbital cycles. Our analysis 
 includes the separated components with a determination of their radial velocity curves, an estimation of the spectroscopic mass ratio, stellar and system
 parameters, and the spectral type of the two components.\\
 To constrain the main stellar parameters we also include in our analysis V-band light curves obtained from the public ASAS archive and 
 VIJK photometric data obtained by Michalska et al. (\cite{michalska}) at the REM 0.6-m telescope at La Silla.\\

\section{Observations}

 To investigate the fundamental properties of DQ Vel we collected a series of high-resolution optical spectra
 between April 2008-2011. Most of the spectra were taken with the 1.2m \emph{Leonard Euler} Swiss telescope 
 at the La Silla Observatory using the CORALIE echelle fiber-fed spectrograph. In our analysis we also included five spectra 
 obtained in June 2008 and May 2009 from the 2.5m \emph{Ir\'{e}n\'{e}e du Pont} telescope echelle spectrograph at the Las Campanas
 Observatory as well as two spectra taken with the echelle spectrograph FEROS mounted at the ESO/MPI 2.2m telescope at
 La Silla. A detailed summary of our spectroscopic observations is given in Table 1. The orbital phases $\phi_{o}$ were calculated using the ephemeris obtained
 by Michalska et al. (\cite{michalska}); they are detailed in Section 3.3.
\begin{table*}
\caption{Detailed optical spectroscopic observations of DQ Vel. $\phi_{o}$ is the orbital phase calculated using the Michalska et al. (\cite{michalska}) 
 ephemeris (see section 3.3). The signal-to-noise ratio was measured using the 5000-5800 $\mathrm{\AA{}}$ range at the continuum level.}     
\centering          
\begin{tabular}{c c c c c c c} 
\hline\hline       
Observatory & Telescope/Instrument & UT date & BJD-2450000 & $\phi_{o}$ & Exp time(s) & S/N \\ 
\hline                    
 La Silla & Euler/CORALIE & 2008-04-05 &  4562.5807 & 0.8061 & 2500 & 53 \\  
 La Silla & Euler/CORALIE & 2008-04-05 &  4562.6105 & 0.8110 & 2500 & 57 \\     
 La Silla & Euler/CORALIE & 2008-04-05 &  4562.6403 & 0.8159 & 2500 & 55 \\  
 La Silla & Euler/CORALIE & 2008-04-07 &  4564.6168 & 0.1408 & 2500 & 60 \\  
 La Silla & Euler/CORALIE & 2008-04-07 &  4564.6466 & 0.1457 & 2500 & 55 \\  
 La Silla & Euler/CORALIE & 2008-04-08 &  4564.6764 & 0.1506 & 2500 & 56 \\  
 La Silla & Euler/CORALIE & 2008-04-08 &  4565.5829 & 0.2996 & 2500 & 63 \\     
 La Silla & Euler/CORALIE & 2008-04-08 &  4565.6127 & 0.3045 & 2500 & 63 \\  
 La Silla & Euler/CORALIE & 2008-04-08 &  4565.6425 & 0.3094 & 2500 & 67 \\  
 La Silla & Euler/CORALIE & 2008-05-22 &  4609.5451 & 0.5263 & 2500 & 58 \\ 
 La Silla & Euler/CORALIE & 2008-05-22 &  4609.5750 & 0.5312 & 2500 & 64 \\  
 La Silla & Euler/CORALIE & 2008-05-22 &  4609.6258 & 0.5395 & 2500 & 57 \\     
 La Silla & Euler/CORALIE & 2008-05-23 &  4610.5343 & 0.6889 & 2500 & 57 \\  
 La Silla & Euler/CORALIE & 2008-05-23 &  4610.5641 & 0.6938 & 2500 & 58 \\  
 La Silla & Euler/CORALIE & 2008-05-23 &  4610.5940 & 0.6987 & 2500 & 55 \\ 
 La Silla & Euler/CORALIE & 2008-05-24 &  4611.5221 & 0.8512 & 2500 & 49 \\  
 La Silla & Euler/CORALIE & 2008-05-24 &  4611.6078 & 0.8653 & 2500 & 46 \\  
 La Silla & Euler/CORALIE & 2008-05-24 &  4611.6377 & 0.8702 & 2500 & 42 \\  
 La Silla & Euler/CORALIE & 2008-05-25 &  4611.6675 & 0.8751 & 2500 & 61 \\  
 La Silla & Euler/CORALIE & 2008-05-25 &  4612.5771 & 0.0247 & 2500 & 34 \\ 
 La Silla & Euler/CORALIE & 2008-05-25 &  4612.6069 & 0.0296 & 2500 & 32 \\  
 La Silla & Euler/CORALIE & 2008-05-25 &  4612.6367 & 0.0345 & 2500 & 31 \\     
 La Silla & Euler/CORALIE & 2008-12-19 &  4819.7745 & 0.0843 & 1500 & 42 \\  
 La Silla & Euler/CORALIE & 2008-12-22 &  4822.7697 & 0.5767 & 1500 & 48 \\  
 La Silla & Euler/CORALIE & 2009-04-15 &  4936.6295 & 0.2932 & 2500 & 62 \\ 
 La Silla & Euler/CORALIE & 2009-04-15 &  4936.6594 & 0.2981 & 2500 & 60 \\  
 La Silla & Euler/CORALIE & 2009-04-15 &  4936.6892 & 0.3030 & 2500 & 59 \\     
 La Silla & Euler/CORALIE & 2009-04-16 &  4937.6087 & 0.4542 & 2500 & 49 \\  
 La Silla & Euler/CORALIE & 2009-04-16 &  4937.6386 & 0.4591 & 2500 & 55 \\  
 La Silla & Euler/CORALIE & 2009-04-16 &  4937.6684 & 0.4640 & 2500 & 53 \\ 
 La Silla & Euler/CORALIE & 2009-04-17 &  4938.6259 & 0.6214 & 2500 & 57 \\  
 La Silla & Euler/CORALIE & 2009-04-17 &  4938.6701 & 0.6287 & 2500 & 54 \\     
 La Silla & Euler/CORALIE & 2009-04-17 &  4938.6999 & 0.6336 & 2500 & 49 \\  
 La Silla & Euler/CORALIE & 2009-05-17 &  4968.5220 & 0.5358 & 2400 & 50 \\  
 La Silla & Euler/CORALIE & 2009-05-17 &  4968.5507 & 0.5405 & 2400 & 47 \\ 
 La Silla & Euler/CORALIE & 2009-05-17 &  4969.5085 & 0.6980 & 2500 & 53 \\  
 La Silla & Euler/CORALIE & 2009-05-18 &  4969.5378 & 0.7028 & 2400 & 55 \\     
 La Silla & Euler/CORALIE & 2009-05-18 &  4970.4967 & 0.8604 & 2400 & 49 \\  
 La Silla & Euler/CORALIE & 2009-05-19 &  4970.5253 & 0.8651 & 2400 & 44 \\  
 La Silla & MPI/FEROS     & 2010-03-28 &  5283.5531 & 0.3214 & 1500 & 67  \\
 La Silla & MPI/FEROS     & 2010-03-30 &  5285.5668 & 0.6524 & 1500 & 62 \\
 Las Campanas & Du Pont/echelle & 2008-06-13  & 4631.4735 & 0.1309 & 900 & 27 \\
 Las Campanas & Du Pont/echelle & 2008-06-13  & 4631.4854 & 0.1329 & 900 & 46 \\
 Las Campanas & Du Pont/echelle & 2009-05-14  & 4966.4639 & 0.1975 & 900 & 60 \\
 Las Campanas & Du Pont/echelle & 2009-05-15  & 4966.5077 & 0.2047 & 900 & 49 \\
 Las Campanas & Du Pont/echelle & 2009-05-16  & 4967.5020 & 0.3681 & 1200& 60 \\
\hline                  
\end{tabular}
\end{table*}
 The echelle spectrograph at Du Pont has a wavelength coverage from $\sim3700$ to 7000 $\mathrm{\AA{}}$. The
 instrument resolution is 45\,000 for one arc-second slit. These data were reduced with
 the standard procedures of de-biasing, flat-fielding, background subtraction, extraction
 of one-dimensional spectra, wavelength calibration, cosmic spike removal and 
 continuum normalisation using the Image Reduction and Analysis Facility IRAF 
 \footnote {IRAF (http://iraf.noao.edu/) is distributed by the National Optical Astronomy Observatories and operated by the Association
 of Universities for Research in Astronomy Inc., under cooperative agreement with the National Science Foundation.}.
 The reduced spectra were then subjected to barycentric corrections.\\
 The high-resolution FEROS spectrograph (R$\sim$48\,000) has a slightly broader spectral range than CORALIE, from 3500 to 9200 $\mathrm{\AA{}}$
 (Kaufer et al. \cite{kaufer1999}, \cite{kaufer2000}).
 The FEROS data presented here were previously reduced by the FEROS Data Reduction System (DRS), 
 which includes barycentric corrections and the standard procedures mentioned above 
 except for continuum normalisation, which was made using the \emph{continuum} task of IRAF.\\
 The CORALIE spectrograph was developed to measure precise stellar radial velocities. 
 The maximum spectral resolution of about 100\,000. The instrument has an automatic reduction software (Baranne et al. \cite{baranne}) 
 that provides fully calibrated images and computes barycentric velocities. Just as with the FEROS spectrograph, CORALIE
 uses a simultaneous Thorium lamp spectrum during the science exposures, which allows
 a regular wavelength calibration. Here we used the CORALIE spectra reduced with this automatic pipeline. Cosmic spikes were removed manually.
 It should be noted here that we found a small-amplitude (lower than 1\% of the total flux) continuum 
 modulation, most likely produced by an imperfect automatic extraction of the orders. 
 To attenuate or remove this modulation we subdivided each CORALIE spectrum into different wavelength intervals. 
 For each interval we applied the \emph{continuum} task using the spline3 function with a corresponding order (different in each interval),
 to create a \textquotedblleft continuum normalised \textquotedblright  interval. We repeated
 this procedure for all spectra using the same wavelength intervals, same function, and same orders as used for the first spectrum.  
 Finally, we combined the corresponding intervals of each spectrum to create a final normalised spectrum.
 Although the data quality significantly improved, there still remain some residuals, which were further attenuated by applying a high-frequency Fourier filter
 to the spectra during the cross-correlation process, as will be explained in Section 3.1.\\
 Our photometric analysis includes a V-band light curve obtained from the public ASAS database \footnote{www.astrouw.edu.pl/asas/}. The observations 
 include 583 frames taken in different epochs during November 2000-2009. Typical errors on these data are less than 0.06 mag.
 The VIJK photometric data used here were obtained with the REM 0.6m telescope at La Silla, which were previously
 reported by Michalska et al. (\cite{michalska}), who decoupled the short- and long-term variabilities using the Fourier decomposition technique.
 The observational campaign included 1051 frames in the V-I bands and 345 frames in the J-K bands. Details of the photometric data are presented in Table 2. \\
\begin{table}
\caption{Detailed photometric data used in this paper. First column is the observing window and $N$ is the number of frames.}                  
\centering                          
\begin{tabular}{c c c c c}      
\hline             
BJD-2450000  & Observatory & Telescope & Filter & N \\
\hline                       
 1998.8357-2885.4690  & La Silla     & REM  & V  & 519 \\
 1998.8372-2899.4789  & La Silla     & REM  & I  & 532 \\
 1998.8360-2865.5494  & La Silla     & REM  & J  & 211 \\
 2000.8472-2855.4982  & La Silla     & REM  & K  & 135 \\
 1868.8363-4971.6065  & Las Campanas & ASAS & V  & 583 \\
\hline                                   
\end{tabular}
\end{table}

\section{Results}
\subsection {Spectral separation and radial velocity curves}

 To investigate if DQ Vel is a single or double line spectroscopic binary we analysed the 46 spectra listed in Table 1.
 We detected known absorption lines and calculated their corresponding Doppler shifts.
 Excluding the intense and broader Balmer lines that seem to contain several components, we selected the stronger and unblended lines:
 $\mathrm{HeI}\,4387.929\,\mathrm{\AA{}}$, $\mathrm{MgII}\,4481.126\,\mathrm{\AA{}}$, $\mathrm{FeI}\,4957.596\,\mathrm{\AA{}}$, 
 $\mathrm{HeI}\,5875.614\,\mathrm{\AA{}}$, the silicon doblet $\mathrm{SiII}\,6347.103\,\mathrm{\AA{}}/6371.359\,\mathrm{\AA{}}$, and
 $\mathrm{HeI}\,6678.151\,\mathrm{\AA{}}$. We directly measured each line position fitting a Gaussian profile using the IRAF task \emph{splot}
 and calculated their corresponding radial velocities. We measured each line at least four times to obtain an estimate of the error.
 We found that all selected HeI lines seem to belong to the more massive star (hereafter gainer) while the metal lines 
 apparently belong to the less massive star (hereafter donor). Averaging the Doppler shifts measured for each line, we calculated
 mean radial velocities (RVs) for the gainer and donor star in each spectrum. The resulting RV measurements are plotted in Figure 1.
  \begin{figure}
   \centering
   \includegraphics[width=9cm,bbllx=50,bblly=10,bburx=490,bbury=350]{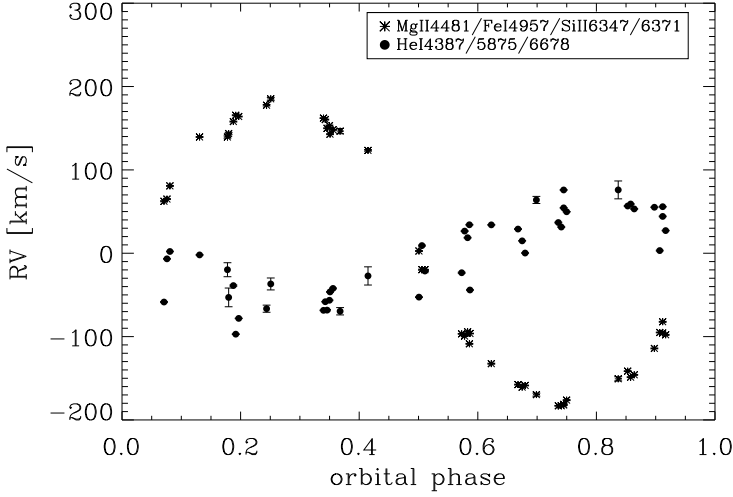}
      \caption{Mean RVs obtained by direct measurement. Donor velocities were obtained averaging the Doppler
       shifts for the $\mathrm{MgII}\,4481\,\mathrm{\AA{}}$, $\mathrm{FeI}\,4957\,\mathrm{\AA{}}$, $\mathrm{SiII}\,6347\,\mathrm{\AA{}}$, and
       $\mathrm{SiII}\,6371\,\mathrm{\AA{}}$ lines. $\mathrm{HeI}$ lines were used to calculate gainer velocities.}
   \end{figure}
Preliminar inspection of these RV curves shows a defined sinusoidal curve for the donor star. The orbit seems to be circular; if eccentricity is present, 
its value is negligible within the errors. \\ 
We observed more scatter on the gainer velocities. We attributed this to the fact that the helium lines are broader than the donor's metal lines, 
which results in a higher 
uncertainty on the determination of the line centre. 
Possible causes for the broadening of the helium lines and/or for the large scatter observed
on the gainer RVs are either a hot rapidly rotating star, as expected for a gainer that is accreting angular momentum together with the material, or
line profiles blended by the contribution of a non-symmetrical pseudo-photosphere or disc.
Yet another cause could be that the gainer is a pulsating star (See section 3.5).
Using the last calculated RVs, we decided to separate the individual spectra to obtain more precise RVs. We used a simple iterative method
originally proposed by Marchenko et al. (\cite{marchenko}), which was used more recently by Gonz\'alez \& Levato (\cite{gonzalez}), who applied this method 
for the orbital analysis of the binary HD143511. 
In general terms, this method is able to separate both components, providing two higher signal-to-noise (S/N) single-line spectra templates,
which can be used to calculate individual RVs, for instance through a cross-correlation process (Griffin \cite{griffin}, Hill \cite{hill}).
The Fourier cross-correlation process (hereafter CCR) convolves two functions (spectra against the template) using a Fourier transform algorithm 
to find the so-called \emph{cross-correlation function (CCF)}. Considering that $\sim$ $80\%$ of our data are from the CORALIE spectrograph and that the accuracy of a 
relative velocity measured with the CCR process increases using a narrow CCF (Hilditch \cite{hilditch}), we applied separating and CCR processes 
only to the CORALIE data. The main reason for this is that lines on the template spectrum and the observed spectra will have similar shapes and widths, 
which decreases the broadening of the CCF.
   \begin{figure}
   \centering
   \includegraphics[width=9cm,bbllx=50,bblly=10,bburx=490,bbury=350]{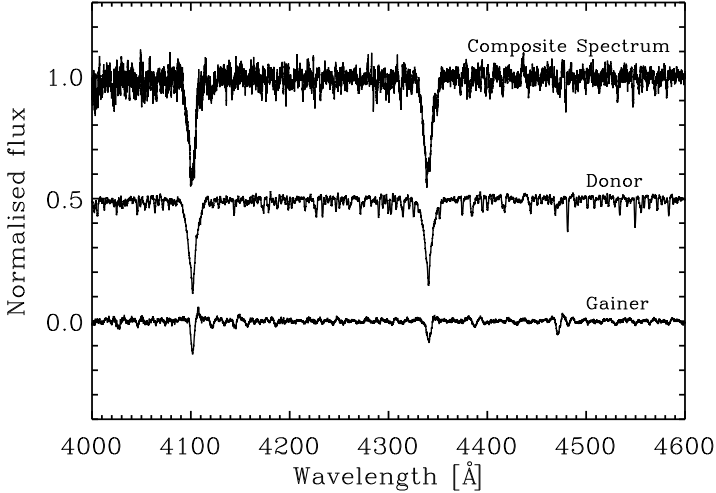}
      \caption{Separated spectra of DQ Vel. The composite spectrum (top) at $\phi_{o}=0.8110$ is used here as a reference from the list in Table 1. 
      Final S/N for the decomposed spectra are 97 (middle) and 267 (bottom) for the donor and gainer star. The spectra were offset for clarity.}
   \end{figure}

   \begin{table}
     \caption[]{Calculated RVs for DQ Velorum using 1D CCR. $RV_\mathrm{d}$ and $RV_\mathrm{g}$ are the mean RV for the donor and gainer star.}
    
  \begin{tabular}{c c r r}
            \hline
            BJD-2450000  & $\phi_{o}$ & $RV_\mathrm{d}\,\mathrm{(km\,s^{-1})}$ & $RV_\mathrm{g}\,\mathrm{(km\,s^{-1})}$\\
            \hline
             4612.5771 & 0.0247 &  $ 62.28\pm0.64$ & $-58.99\pm0.63 $  \\
             4612.6069 & 0.0296 &  $ 66.37\pm0.59$ &  $-6.90\pm0.49$    \\
             4612.6367 & 0.0345 &  $ 81.65\pm0.46$ &   $2.12\pm0.32$    \\
             4819.7745 & 0.0843 &  $139.21\pm0.60$ &  $-2.74\pm0.42$   \\
             4564.6168 & 0.1408 &  $157.54\pm0.69$ & $-57.80\pm2.48$    \\
             4564.6466 & 0.1457 &  $163.68\pm0.53$ & $-97.07\pm0.42$   \\
             4564.6764 & 0.1506 &  $164.80\pm0.52$ & $-78.05\pm0.30$   \\
             4936.6295 & 0.2932 &  $161.90\pm0.42$ & $-68.49\pm0.41$    \\
             4936.6594 & 0.2981 &  $160.04\pm0.61$ & $-57.92\pm0.42$    \\
             4565.5829 & 0.2996 &  $150.65\pm0.43$ & $-67.78\pm1.03$    \\
             4936.6892 & 0.3030 &  $153.33\pm0.62$ & $-56.34\pm0.53$    \\
             4565.6127 & 0.3045 &  $143.18\pm0.52$ & $-46.62\pm0.54$   \\
             4565.6425 & 0.3094 &  $146.92\pm0.54$ & $-41.47\pm0.89$  \\
             4937.6087 & 0.4542 &  $  3.89\pm0.54$ & $-52.49\pm0.42$    \\
             4937.6386 & 0.4591 & $ -17.44\pm0.77$ & $  9.33\pm0.57$   \\
             4937.6684 & 0.4640 & $ -18.72\pm0.53$ & $-21.33\pm0.37$    \\
             4609.5451 & 0.5263 & $ -95.82\pm0.50$ & $-23.60\pm0.73$    \\
             4609.5750 & 0.5312 & $ -99.43\pm0.52$ & $ 26.70\pm0.52$   \\
             4968.5220 & 0.5358 & $ -95.47\pm0.54$ & $ 18.83\pm0.48$   \\
             4609.6258 & 0.5395 & $-108.24\pm0.51$ & $ 34.00\pm0.44$   \\
             4968.5507 & 0.5405 &  $-97.11\pm0.57$ & $-44.20\pm0.33$    \\
             4822.7697 & 0.5767 & $-131.58\pm0.58$ & $ 34.14\pm0.37$    \\
             4938.6259 & 0.6214 & $-157.34\pm0.52$ & $ 29.28\pm0.45$    \\
             4938.6701 & 0.6287 & $-160.61\pm0.49$ & $ 15.12\pm0.27$    \\
             4938.6999 & 0.6336 & $-159.57\pm0.49$ & $  0.25\pm0.24$    \\
             4610.5343 & 0.6889 & $-184.89\pm0.46$ & $ 37.11\pm0.59$    \\
             4610.5641 & 0.6938 & $-183.61\pm0.51$ & $ 31.42\pm0.43$    \\
             4610.5940 & 0.6987 & $-183.13\pm0.50$ & $ 54.30\pm0.38$   \\
             4969.5085 & 0.6980 & $-180.27\pm0.64$ & $ 75.60\pm0.37$     \\
             4969.5378 & 0.7028 & $-177.57\pm0.50$ & $ 49.58\pm0.34$     \\
             4562.5807 & 0.8061 & $-141.67\pm0.44$ & $ 56.73\pm0.42$   \\
             4562.6105 & 0.8110 & $-146.29\pm0.49$ & $ 58.45\pm0.38$   \\
             4562.6403 & 0.8159 & $-144.55\pm0.47$ & $ 53.10\pm0.37$   \\
             4611.5221 & 0.8512 & $-112.38\pm0.57$ &  $55.44\pm0.39$   \\
             4970.4967 & 0.8604 &  $-96.28\pm0.41$ &   $3.68\pm0.49$      \\
             4970.5253 & 0.8651 &  $-84.37\pm0.59$ &  $55.73\pm0.37$       \\
             4611.6078 & 0.8653 &  $-96.31\pm0.61$ &  $44.25\pm0.49 $ \\
             4611.6377 & 0.8702 & $-97.31\pm0.59$  &  $27.13\pm0.37$   \\
            \hline
\end{tabular}
   \end{table}
Using previously calculated CORALIE RVs, we shifted all spectra to the donor rest-frame and 
combined these to create a first-guess spectral template for the donor star. This template is an orbital-cycle averaged single-line spectrum.
This procedure attenuates the spectral features of the remaining component (gainer).
Subtracting this donor template from all spectra and shifting the resulting spectra to the gainer rest-frame, we
combined the data again to obtain an orbital-cycle-averaged template for the gainer star. We repeated this procedure until the residuals
of one component contribution disappeared from the template spectra of the other. According to Gonz\'alez \& Levato (\cite{gonzalez}), 
no more than 5-7 iterations are necessary considering that with each iteration the residuals are reduced by a  
factor $1/n$, where $n$ is the number of observed spectra. Figure 2 shows a portion of our separated spectra after six iterations. \\
We applied a 1D Fourier CCR process to the observed spectra and previously obtained templates using the IRAF \emph{fxcor} task. 
For binaries this task works as follows: we select the template of one component (which is by construction at rest in the barycentric frame), and convolve it
 with each of the observed spectra in which the
spectral lines move according to the relative motion of the binary. The maximum of the CCF will be, in this case, the shifted velocity of the spectrum relative to the
template. The process is then repeated using the other template. In this way, we obtain the relative velocities of each spectrum with respect to the donor and gainer
templates.\\
During the CCR process and to finally remove the weak modulation found on the CORALIE spectra, we applied a Hanning-type filter to the template and
the observed spectra after the data were transformed and prior to correlation. This Hanning-filter function is used to attenuate the Fourier components over
a specific wavenumber range to remove remaining high-frequency noise that was not removed by the continuum subtraction.\\
From the donor template, we calculated RVs independently in two spectral ranges at 4500-4800\,$\mathrm{\AA{}}$ and 4900-5600\,$\mathrm{\AA{}}$, where we found more
spectral lines from the donor star.
A third measurement was made in both spectral ranges simultaneously. Final RVs were obtained by taking the mean between these three measurements.
We applied the same procedure using the gainer template in the 4350-4750\,$\mathrm{\AA{}}$ and 4900-5300\,$\mathrm{\AA{}}$ ranges. 
Table 3 shows our final RVs determined from the CCR process and their corresponding r.m.s. errors.\\
In Table 4 we show the RVs for the remaining spectra that were not included in the CCR process.\\
   \begin{table}
     \caption{RVs for DuPont/echelle and MPI/FEROS spectra not included in the CCR process.}
\centering
        \begin{tabular}{c c r r}
            \hline
            BJD-2450000  & $\phi_{o}$ & $RV_\mathrm{d} \mathrm{(km\,s^{-1})}$ & $RV_\mathrm{g} \mathrm{(km\,s^{-1})}$\\
            \hline
             4631.4735 & 0.1309 & $ 139.40\pm1.33$ & $-19.78\pm8.38 $   \\
             4631.4854 & 0.1329 &  $143.62\pm1.37$ & $-53.00\pm11.23$   \\
             4966.4639 & 0.1975 &  $177.57\pm0.89$ & $-66.54\pm4.28 $   \\
             4966.5077 & 0.2047 &  $185.23\pm1.72$ & $-36.84\pm7.14 $   \\
             5283.5531 & 0.3214 &  $146.55\pm3.13$ & $-69.58\pm4.41 $   \\
             4967.5020 & 0.3681 & $ 123.41\pm1.00$ & $-27.20\pm10.91$   \\
             5285.5668 & 0.6524 & $-169.66\pm1.79$ & $ 63.94\pm4.26 $   \\
            \hline
          \end{tabular}
   \end{table}
To use all available RV measurements, we decided to obtain the mean RV value in orbital phase bins of 0.05.
We fitted the RV curves with a sinus function. Best fits and circular-orbit solutions are shown in Figure 3 and Table 5.
The different $\gamma$ obtained for the gainer suggests that their lines are influenced by circumstellar material.
 \begin{figure}
   \centering
   \includegraphics[width=9cm,bbllx=50,bblly=10,bburx=490,bbury=350]{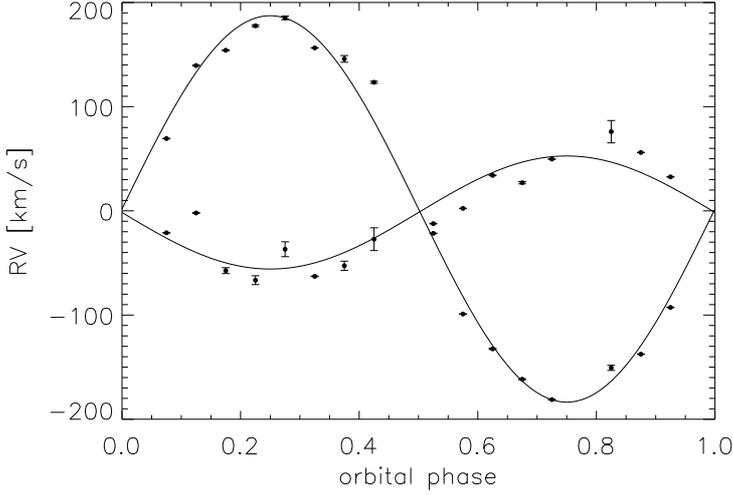}
      \caption{RV curves and best sinus fit for DQ Vel components. Each RV was measured from Table 3 and Table 4 taking the mean value in orbital
      phase bins of 0.05.}
   \end{figure}

   \begin{table}
     \caption{Circular-orbit solutions obtained for the best sinus fits. }
\centering
\begin{tabular}{l c c}
            \hline
            Parameter  & Donor star & Gainer star \\
            \hline
                K ($\mathrm{km\,s^{-1}}$)   & $185.436\pm4.963$ & $ -57.556\pm5.536 $ \\
            $ \gamma$ ($\mathrm{km\,s^{-1}}$) & $ 1.858\pm3.712$  & $ -8.745\pm4.066 $ \\
              r.m.s ($\mathrm{km\,s^{-1}}$) & 14.80 &  14.07    \\
            \hline
          \end{tabular}
\tablefoot{$K$ is the semi-amplitude of RV curves and $\gamma$ is the systemic velocity. Data included here were obtained from the CCR process and direct measurements.}
   \end{table}

Based on these results, we determined the projected semi-major axes and minimum masses for donor and gainer star using the expressions by Hilditch 
(\cite{hilditch}) for a circular orbit:
\begin{equation}
 a_\mathrm{d,g} \sin i = (1.9758 \times 10^{-2}) K_\mathrm{d,g} P_\mathrm{o}\, [R_{\odot}]
\end{equation}
\begin{equation}
  a_\mathrm{d} \sin i = 22.288 \pm 0.596 \, [R_{\odot}]
\end{equation}
\begin{equation}
  a_\mathrm{g} \sin i = 6.917 \pm 0.665 \, [R_{\odot}]
\end{equation}
\begin{equation}
 m_\mathrm{d,g} \sin^{3} i = (1.0361 \times 10^{-7})(K_\mathrm{d}+K_\mathrm{g})^{2} K_\mathrm{g,d} P_\mathrm{o}\, [M_{\odot}]
\end{equation}
\begin{equation}
  m_\mathrm{d} \sin^{3} i = 2.141 \pm 0.244\, [M_{\odot}]
\end{equation}
\begin{equation}
  m_\mathrm{g} \sin^{3} i = 6.901 \pm 0.460\, [M_{\odot}]\,,
\end{equation}
where  $a_\mathrm{d}$ and $a_\mathrm{g}$ are the donor and gainer semi-major axes measured in \emph{solar radii} ($R_{\odot}$), $m_\mathrm{d}$ and $m_\mathrm{g}$ are the 
donor and gainer masses measured in \emph{solar masses} ($M_{\odot}$), $K_\mathrm{d}$ and $K_\mathrm{g}$ are the semi-amplitudes of the RV curves expressed in $\mathrm{km\,s^{-1}}$,
$P_\mathrm{o}$ is the orbital period in days, and $i$ the orbital inclination of the system. \\
It can be shown that independently of the orbital inclination, equation 5 yields
\begin{equation}
 \frac{m_\mathrm{d}}{m_\mathrm{g}}=\frac{K_\mathrm{g}}{K_\mathrm{d}}=q=0.31\pm0.03\,,
\end{equation}
which represents the spectroscopic mass ratio $q$ of the system. Note that for this $q$ value we assumed that the observed RV variations of the
gainer's lines represent the motion of the gainer.\\
If we consider that the donor star fills its Roche lobe, we can approximate the donor radius ($R_\mathrm{d}$) to the size of the donor Roche lobe according to Eggleton 
(\cite{eggleton}):
\begin{equation}
 r_\mathrm{L,d}=\frac{R_\mathrm{d}}{a}=\frac{0.49q^{2/3}}{0.69q^{2/3}+\ln(1+q^{1/3})}\,,
\end{equation}
where $a$ is the semimajor axis of the relative orbit.\\
For $q=0.31\pm0.03$, we obtain
\begin{equation}
 R_\mathrm{d}= (0.269 \pm 0.019)\,a\,.
\end{equation}
For an eclipsing binary with an orbital inclination $70^\circ \leq i \leq 90^\circ$ we can estimate a donor radius between 
$(7.856 \pm 0.604)\, R_{\odot} \leq R_\mathrm{d} \leq (8.366\pm0.643)\, R_{\odot}$, which indicates an oversized star based on the previous mass value.\\
We can also determine the maximum value for the equatorial rotational velocity of the donor assuming it is rotating synchronously:
\begin{equation}
 v_\mathrm{r,d}=\frac{2\pi R_\mathrm{d}}{P_\mathrm{o}}=69.606\pm5.349 \, \mathrm{km\,s^{-1}}\,,
\end{equation}
for which we used the maximum donor radius obtained above.

\subsection {Spectral type classification and analysis of spectral features}
 
 The donor star template shows several absorption metal lines including $\mathrm{FeI-II}$, $\mathrm{MgI-II}$, $\mathrm{TiII}$, $\mathrm{CrII}$, and 
$\mathrm{SiII}$. 
 Some lines like $\mathrm{FeII}\,4233.2\,\mathrm{\AA{}}$, $\mathrm{FeI}\,4383.5\,\mathrm{\AA{}}$, $\mathrm{MgI}\,5167.3\,\mathrm{\AA{}}$, 
 $\mathrm{MgII}\,5528.4\,\mathrm{\AA{}}$, and $\mathrm{SiII}\,5688.8\,\mathrm{\AA{}}$ were not clearly identified in the original spectra. 
 We compared our donor template with a grid of synthetic spectra in the region 5200-5350\,$\mathrm{\AA{}}$ which is deprived of hydrogen and helium lines but 
has several metal lines.
 To determine the grid of synthetic fluxes we used atmospheric models computed with the line-blanketed LTE ATLAS9 code (Kurucz \cite{kurucz}), which treats line 
 opacity with opacity distribution functions (ODFs). The Kurucz models are constructed assuming plane-parallel geometry and hydrostatic 
 and radiative equilibrium of the gas. The synthetic spectra were computed with the SYNTHE code (Kurucz 1993). Both codes, ATLAS9 and SYNTHE, were ported
 under GNU Linux by Sbordone (\cite{sbordone}) and are available online \footnote{wwwuser.oat.ts.astro.it/atmos/}. The atomic data were taken from 
 Castelli \& Hubrig \cite{castelli} \footnote{http://wwwuser.oat.ts.astro.it/castelli/grids.html}. The theoretical models were obtained for effective 
 temperatures from 6000 to 10\,000 $\mathrm{K}$ with steps of 100 $\mathrm{K}$ and for surface gravities from 2.0 to 4.5 dex with steps of 0.1 dex. Solar and 0.5 dex 
 higher metallicities were taken into account. The grid of synthetic spectra was calculated for five different rotation velocities, $v_\mathrm{r} \sin i =$ 
 0, 25, 50, 75 and 100 $\mathrm{km\,s^{-1}}$.
 We subtracted the donor template from every grid spectrum and analysed the residuals of the resulting spectra.
 The best model is obtained for an A-type star with $T_\mathrm{d}= 9400\ \mathrm{K}$, $\log g_\mathrm{d}$ = 3.1, $v_\mathrm{r}\ \sin i = 75 \mathrm{km\,s^{-1}}$ and solar metallicity.
To obtain a more restricted value for the rotational velocity, we decided to interpolate between the different values of rotation velocities
generated in the grid of synthetic spectra. We selected four synthetic spectra from the grid with 
$T = 9400\ \mathrm{K}$, $\log g = 3.1$ and velocities of
 25, 50, 75, and 100 $\mathrm{km\,s^{-1}}$. We measured the full width at half maximum (\emph{fwhm}) for the isolated and unblended metal lines 
$\mathrm{MgI}\,4703.0\,\mathrm{\AA{}}$, $\mathrm{MgII}\,4481.3\,\mathrm{\AA{}}$, $\mathrm{TiII}\,4780.0/4805.1\,\mathrm{\AA{}}$, and $\mathrm{SiII}\,6371.3\,\mathrm{\AA{}}$. 
We plot in Figure 4 the \emph{fwhm}-$v_\mathrm{r} \sin i$ calibration curves.
 \begin{figure}
   \centering
   \includegraphics[width=8.8cm,bbllx=50,bblly=10,bburx=490,bbury=350]{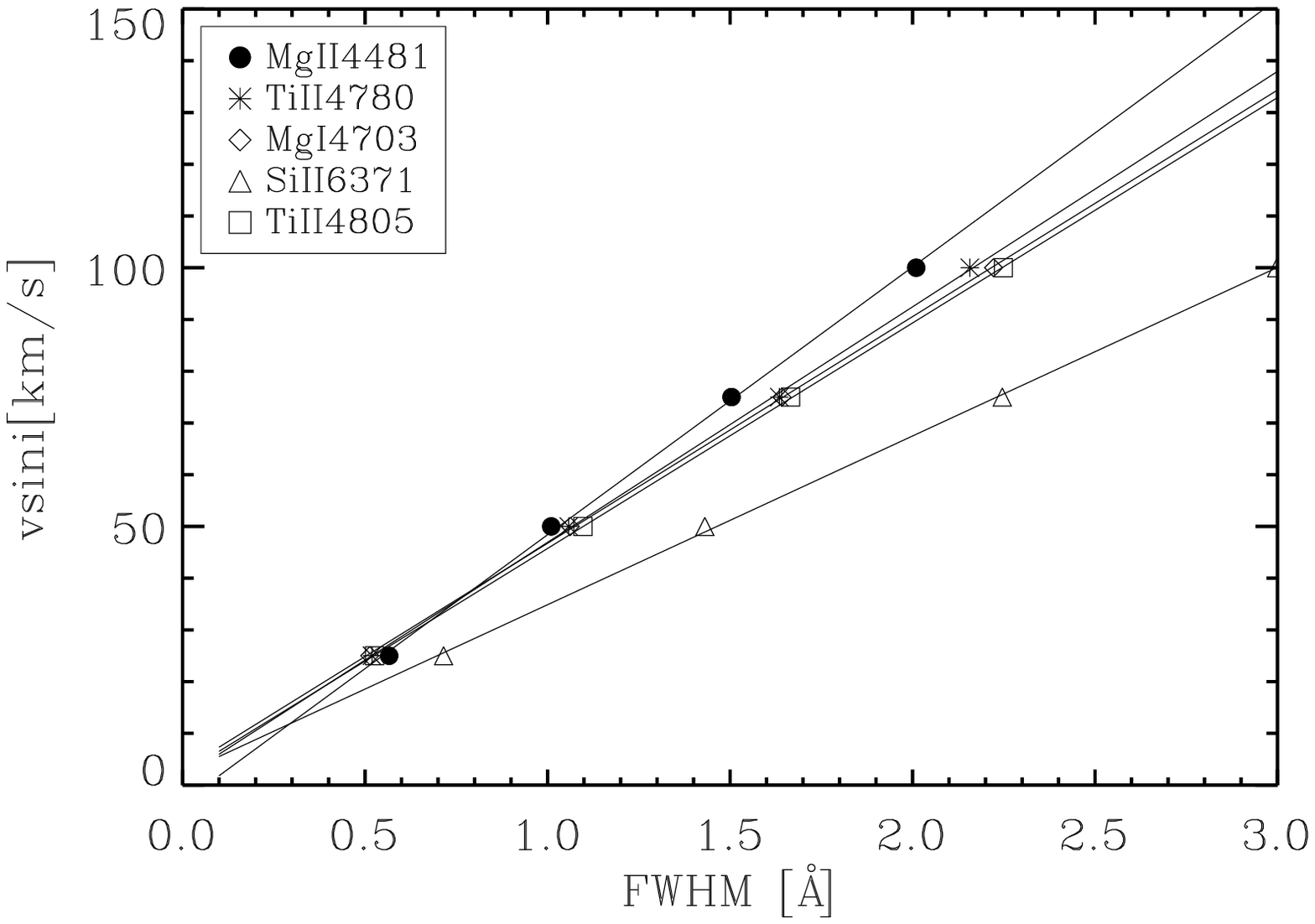}
      \caption{Calibration curves for measured fwhm ($\mathrm{\AA{}}$) of $\mathrm{MgI}\,4703.0\,\mathrm{\AA{}}$, $\mathrm{MgII}\,4481.3\,\mathrm{\AA{}}$, $\mathrm{TiII}\,4780.0/4805.1\,\mathrm{\AA{}}$, 
 and $\mathrm{SiII}\,6371.3\,\mathrm{\AA{}}$ lines plotted against $v_\mathrm{r}\,\sin\,i$.} 
   \end{figure}
We calculated the \emph{fwhm} for the same spectral lines in the donor template and determined a mean velocity of 
$v_\mathrm{r,d} \sin i = 65\pm4$ $\mathrm{km\,s^{-1}}$, which agrees with our previous result of equation 11.\\
A first inspection of spectral features in the gainer template suggests a B-type star with marked Balmer and helium lines 
like $\mathrm{HeI}\,4026/\,4471/\,4921/\,5015/\,5875/\,6678\,\mathrm{\AA{}}$ and also a few and weak metal lines such as 
$\mathrm{FeII}\,4303/\,5169/\,5276\,\mathrm{\AA{}}$, $\mathrm{MgII}\,4481\,\AA{}$, and $\mathrm{SiII}\,4133\,\mathrm{\AA{}}$.
We also observed asymmetric profiles in the helium and Balmer lines after the separation, as shown in Figure 7.\\
In the current DPV scenario, the donor star fills its Roche lobe and transfers mass to a rapidly rotating gainer star via gas stream. 
Therefore, the analysis of the gainer spectra requieres special attention to check for the possible presence of circumstellar material. To study the
spectral-orbital behaviour of the gainer, we decided to remove the donor contribution from the original spectra using the previously found donor template. 
To derive the contribution of the donor to the total light along the orbital phase, we modeled the light curve in the V-band with the previously calculated 
donor temperature and mass ratio and assumed that the donor light adds to the total light. Under these assumptions, the total light includes two stellar 
components (L1 and L2) plus a circumstellar disc component (Ld). More details about the photometric model are explained in section 3.4.
We calculated the donor factor contribution to the total light for specific spectral lines. These factors were determined taking into account the orbital changes
in the proyected area of the donor as well as the wavelength dependence on the different colour terms of donor and gainer (see Fig.5).
\begin{figure}
   \centering
   \includegraphics[width=8cm]{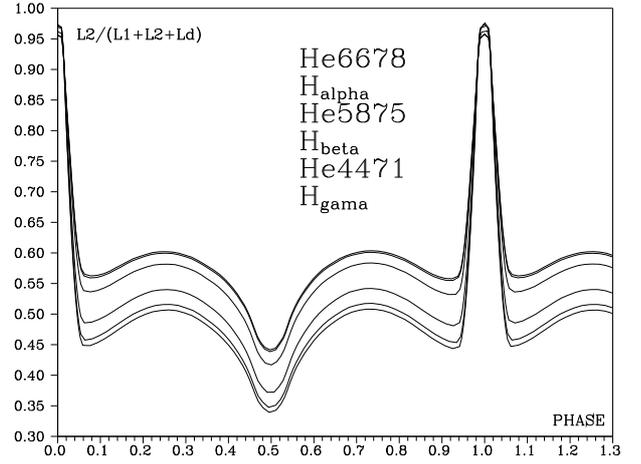}
      \caption{Donor light contribution with respect to the total light calculated from the light curve model. Each curve shows the donor contribution for a specific 
spectral range around a spectral line. The selected spectral lines (and ranges) are from top to the bottom: $\mathrm{HeI}\,6678\,\mathrm{\AA{}}$, 
$\mathrm{H\alpha}$, $\mathrm{HeI}\,5875\,\mathrm{\AA{}}$, $\mathrm{H\beta}$, 
$\mathrm{HeI}\,4471\,\mathrm{\AA{}}$, and $\mathrm{H\gamma}$. L1 and L2 are the stellar fluxes and Ld is the disc contribution.} 
   \end{figure}
From each observed spectrum, we subtracted the donor template multiplied by the corresponding factor for the specific wavelength range.
The resulting spectra were normalised to the new continuum level. Selected donor-subtracted $\mathrm{H\alpha}$ and $\mathrm{H\beta}$ profiles are shown 
in Figure 6.
\begin{figure}
   \centering
   \includegraphics[width=9cm]{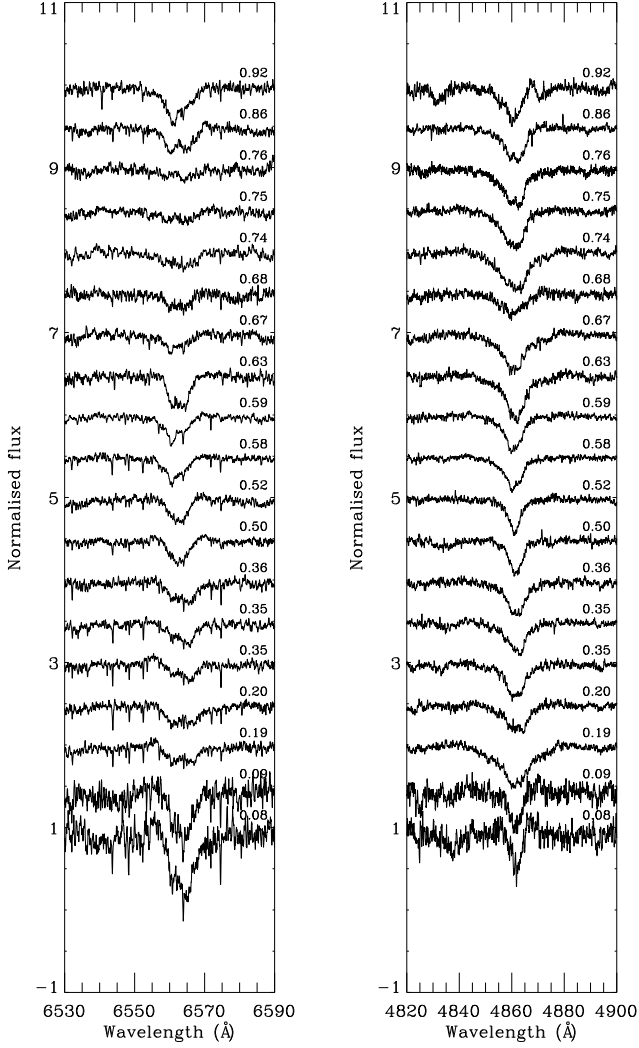}
      \caption{Selected donor-subtracted $\mathrm{H\alpha}$ (left) and $\mathrm{H\beta}$ (right) profiles along the orbital period. 
        Orbital phases are shown on the right side of the left plot. Same values were used in the right panel.} 
   \end{figure}
The two profiles are composed of a central absorption surrounded by weak blue/red asymmetric wings, whose velocities reach up to about $\sim$ 350 $\mathrm{km\,s^{-1}}$
and are phased with the gainer's RVs. Similar behaviour was found in $\mathrm {H\gamma}$ profiles. 
We detected a weakness of central absorption during the orbital phases $\phi_\mathrm{o}=0.1-0.2$ and $\phi_\mathrm{o}=0.6-0.7$ also observed in the helium lines
and probably associated to a filling emission from a higher local temperature interaction regions such as hot or bright spots, visible during those orbital
phases.
\begin{figure}
   \centering
  \includegraphics[width=9cm]{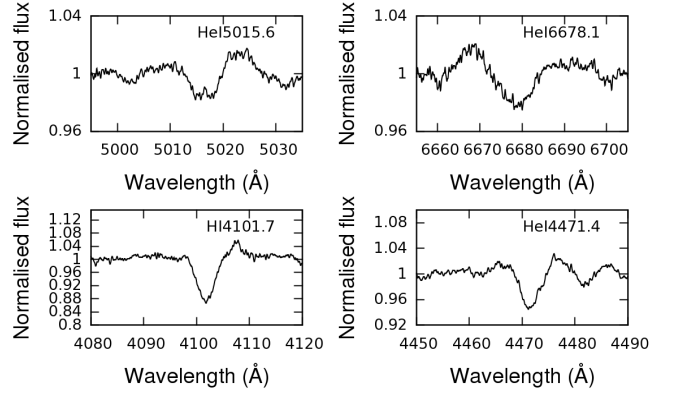}
      \caption{Asymmetric helium and Balmer profiles found in the gainer template after separating the spectra.} 
   \end{figure}
\\
To gain some insight into the gainer temperature we calculated equivalent widths (hereafter EW) for the $\mathrm{HeI}\,4471\,\mathrm{\AA{}}$ and $\mathrm{MgII}\,4481\,\mathrm{\AA{}}$ lines
for all donor-subtracted spectra. Mennickent at al. (\cite{mennickent2012b}) determined the EWs ratios $R=EW_\mathrm{4471}/EW_\mathrm{4481}$ for B-type stars 
using a grid of synthetic spectra ($\log g=4.0$, $v_\mathrm{turb}=2 \mathrm{km\,s^{-1}}$ and solar metallicity) and plotted it against temperatures.
These results were used to determine the temperature of the gainer. Similarly, we calculated $R$ in all donor-subtracted spectra and used the mean value to 
estimate a temperature $T_\mathrm{g}\sim(15\,000\pm1000)\ \mathrm{K}$ for the gainer star. We show the variability of $EW_\mathrm{4471}$ and $R$ along the orbital phase in Figure 8.\\
\begin{figure}
   \centering
   \includegraphics[width=9cm]{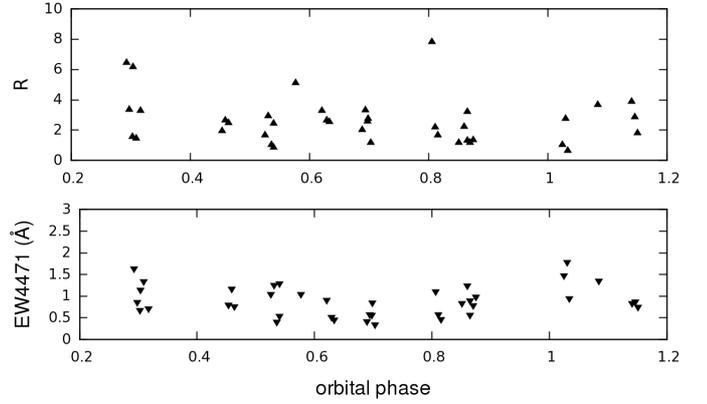}
      \caption{$EW_\mathrm{4471}$ (bottom) and $R=EW_\mathrm{4471}/EW_\mathrm{4481}$ (top) plotted against orbital phase.}
   \end{figure}

\subsection{Orbital and long light curves}
 We separated multiperiodic V, I, J, and K light curves with a Fourier decomposition technique (for more details see Demircan \cite{demircan}, Mennickent et al. 
\cite{mennickent2012a})
and found two main Fourier components of frequencies $f_{1}$ and $f_{2}$, the sum of which (including their most important harmonics) is the
best representation of the light curves.
$f_{1}$ is higher and associated with a period of 6.083299 d, which represents the orbital variability of the binary (see Fig.9).
We find the following orbital cycle ephemeris for the light curve minimum:
\begin{equation}
 HJD_\mathrm{min,orb}= 2453407.60(2)+6.083299(7)\times E\,,
\end{equation}
which agrees with the ephemeris from van Houten (\cite{vanhouten}).\\
 \begin{figure}
   \centering
   \includegraphics[width=9cm]{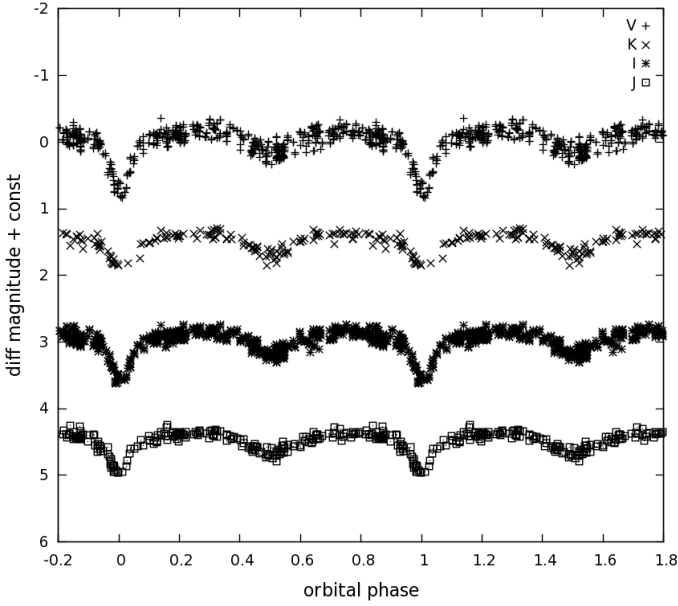}
      \caption{Separated VIJK orbital light curves during one season in 2008 phased with the orbital period.}
   \end{figure}
The frequency $f_{2}$ is related to a period of 188.7 d and is represented by a smooth variability whose amplitude in V-band is $\sim 2\%$ with 
respect to the total luminosity
(see Fig.10).
The ephemeris for the maximum of the long-term variability is best represent by
\begin{equation}
 HJD_\mathrm{max,long}= 2453437.2(16)+188.7(2)\times E\,.
\end{equation}
 \begin{figure}
   \centering
   \includegraphics[width=8cm]{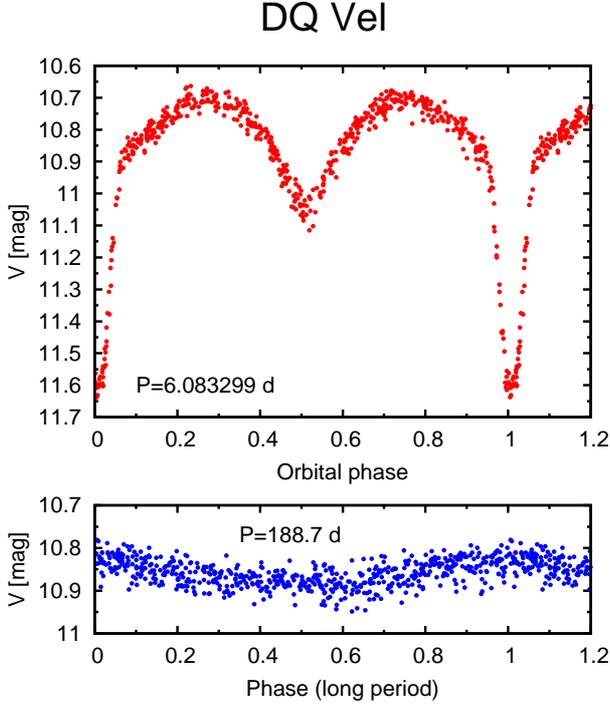}
      \caption{DQ Vel ASAS V-band light curve after separating the short-term (top) and long-term (bottom) variability.}
   \end{figure}
A visual inspection of the orbital light curves shows an eclipsing variable with well-defined eclipses and noticeable proximity effects. Clearly,
at least one of the components is tidally distorted. Differences in the depths of the minima suggest different surfaces temperatures.
The deepest primary minimum on the V-band indicates that the bluer, hotter, and smaller gainer star is being eclipsed. Its light contribution is even less dominant on the
infrared bands, supporting our previous results from the spectroscopic studies.\\

\subsection{Light curve modelling}

To derive the main physical parameters for the stellar components we fitted the V-band light curve with the inverse-problem-solving 
method based on the simplex algorithm. Main binary elements used in the model and the light-curve-fitting procedure are described in 
Djura\v{s}evi\'{c} (\cite{djurasevic1992}, \cite{djurasevic1996}).
We assumed a semi-detached configuration with the donor filling its Roche lobe. After unsuccessful attempts to model the system without a disc, we finally
adopted a configuration that included an optically thick accretion disc around the gainer star.
The disc is characterised by its radially dependent temperature $T_\mathrm{disc}(r)$, 
its radius $R_\mathrm{disc}$, its thickness at edge $d_\mathrm{e}$ and at centre $d_\mathrm{c}$, and by assuming that it is in physical and thermal contact with the gainer. 
The temperature distribution of the disc is given by
\begin{equation}
 T_\mathrm{disc}(r)= T_\mathrm{disc_{e}}+(T_\mathrm{g}-T_\mathrm{disc_{e}})\left[1-\frac{r-R_\mathrm{g}}{R_\mathrm{disc}-R_\mathrm{g}}\right]^{a_\mathrm{T}}\,,
\end{equation}
where $T_\mathrm{disc_{e}}$ is the temperature of the disc at the edge, $a^\mathrm{T}$ is the temperature exponent, and ($T_\mathrm{g}, R_\mathrm{g}$) 
represent the temperature and radius of the gainer star.\\
The model also includes two active regions, \emph{hot spot} (hs) and \emph{bright spot} (bs), with higher local temperatures, which are located at the edge of the disc.
Including these regions significantly improved the fit. They are represented by their temperatures ($T_\mathrm{hs,bs}$), angular dimensions ($\theta_\mathrm{hs,bs}$), and
longitudes ($\lambda_\mathrm{hs,bs}$), which are free
parameters in the model. The longitudes $\lambda_\mathrm{hs,bs}$ are measured clockwise in the range $0-360$ degrees and are viewed from the direction of the +z-axis, which 
is orthogonal to the orbital plane. More details about the disc model are described in Mennickent et al. \cite{mennickent2012a}.\\
To restrict the number of free parameteres we fixed the donor temperature $T_\mathrm{d}=9400\ \mathrm{K}$ and mass ratio $q=0.31$, whose values were reliably derived from
the previous spectroscopic analysis (Sect.3.1 and 3.2).
We also fixed the gravity-darkening and albedo coefficients of the components to the values $\beta_\mathrm{d,g}=0.25$ and 
$A_\mathrm{d,g}=1.0$ according to the von Zeipel law for radiative shells (von Zeipel \cite{vonzeipel}). 
Considering that the donor star fills its Roche lobe, we assumed this star to rotate synchronously with a rotation coefficient $f_\mathrm{d}=1.0$. 
We modeled the ASAS V-band light curve with the gainer star in synchronous-rotation regime ($f_\mathrm{g}=1.0$) and also in a critical-rotation scenario
with a non-synchronous rotation coefficient $f_\mathrm{g}=12.8$. 
We observed no significant difference in the two fitting results, revealing a minor effect of the gainer rotational velocity on the system parameters.
The V-band light curve fitting with the gainer in synchronous rotation is shown in Figure 11. Table 6 shows the final parameters of the fitting. 
The main physical properties of the gainer star such as mass ($M_\mathrm{g}=7.3\pm0.3\ M_{\odot}$), radius ($3.6\pm0.2\ R_{\odot}$), and gravity 
($\log g_\mathrm{g}=4.2\pm0.1$) agree with our previous spectroscopic result for the spectral classification (B-type star).
However, we find a temperature value for the gainer $T_\mathrm {g}=18\,500\pm500\ \mathrm{K}$ different from the previously obtained $T_\mathrm{g}=15\,000\pm1000\mathrm{K}$ using the equivalent
width ratio between the spectral features $R=EW_\mathrm{4471}/EW_\mathrm{4481}$. This difference is probably associated with the disc contribution to the spectra.\\
As was mentioned before, the best-fit model for DQ Vel includes an optically thick disc around the gainer star.
From Figure 11, we estimate a disc contribution of about $7\%$ of the total light at quadrature phases. It has a concave shape with higher thickness at its
edge than at its centre and is extended. It fills $(89\pm3)\%$ of the critical gainer Roche lobe radius with a temperature of $6580\pm300\ \mathrm{K}$ at the outer radius,
 as shown in Table 6.
According to the model, the hot spot --where the material coming from the donor impacts the disc-- is located at longitude $329.1^{\circ}\pm7.0^{\circ}$ and has a 
temperature of $T_\mathrm{hs}= 9500\pm750 \mathrm{K}$, which is close to the donor temperature found previously. This is probably the reason why we do not
 observe emission from the hot spot on the
composite spectra. Because the disc is extended, the distance between $L1$ and the outer radius of the disc is short and not sufficient to produce a substantial 
acceleration of the gas stream and thus reach a higher temperature at this region.
\begin{table*}
\caption{Analysis results of the DQ Vel V-band light-curve.}
\centering
        \begin{tabular}[b]{l l l}
            \hline
{Quantity} &  & Remarks \\
            \hline
 
  $n$                               &  583                      & Number of observations.\\
 $\mathrm{\Sigma}(O-C)^2$           &  0.3984                   & Final sum of the residuals square between observed (LCO) and synthetic (LCC) light-curves.\\
$\mathrm{\sigma_{rms}}$             & 0.0258                    & Root-mean-square of the residuals.\\
$ i^\mathrm{\circ} $                & $82.5\pm0.2$              & Orbit inclination (in arc degrees).\\
$F_\mathrm{disc}=R_\mathrm{disc}/R_\mathrm{yc}$ & $0.89\pm0.03$ & Disc dimension factor (the ratio of the disc radius to the critical Roche lobe radius along the y-axis).\\
$T_\mathrm{disc} [\mathrm{K}]$      & $6580\pm300$              & Disc-edge temperature.\\
$d_\mathrm{e} [a_\mathrm{orb}]$     & $0.019\pm 0.005$          & Disc thicknesses at the edge in the units of the distance between the components.\\
$d_\mathrm{c} [a_\mathrm{orb}]$     & $0.011\pm 0.005$          & Disc thicknesses at the centre in the units of the distance between the components. \\
$a_\mathrm{T}$                      & $4.5\pm 0.3$              & Disc temperature distribution coefficient.  \\
$F_\mathrm{g}$                      & $0.271\pm 0.005$          & Filling factor for the critical Roche lobe of the gainer in synchronous rotation regime.\\
$T_\mathrm{g} [\mathrm{K}]$         & $18\,500\pm 500$          & Temperature of the gainer.\\
$A_\mathrm{hs}=T_\mathrm{hs}/T_\mathrm{disc}$   & $1.45\pm0.1$  & Hot-spot temperature coefficient. \\
$\mathrm{\theta_{hs}}[^\mathrm{\circ}]$  & $16.0\pm 2.0$        & Hot-spot angular dimension.\\
$\mathrm{\lambda_{hs}} [^\mathrm{\circ}]$ & $329.1\pm7.0$       & Hot-spot longitude (in arc degrees). \\
$\mathrm{\theta_{rad}} [^\mathrm{\circ}]$ & $-19.7 \pm 5.0$     & Angle between the line perpendicular to the disc edge and the direction of the hot spot.\\
$A_\mathrm{bs}=T_\mathrm{bs}/T_\mathrm{disc}$  & $1.39\pm 0.1$  & Bright-spot temperature coefficient. \\
$\mathrm{\theta_{bs}} [^\mathrm{\circ}]$ & $50.1\pm 5.0$        & Bright-spot angular dimension.\\
$\mathrm{\lambda_{bs}} [^\mathrm{\circ}]$ & $142.7\pm 9.0$      & Bright-spot longitude (in arc degrees).\\
$\mathrm{\Omega_{g}}$              & $8.45\pm 0.02$             & Dimensionless surface potentials of the gainer.\\
$\mathrm{\Omega_{d}}$          & $2.49\pm 0.02$                 & Dimensionless surface potentials of the donor.\\
$M_\mathrm{g}\ [M_{\odot}]$ & $7.3\pm0.3$                       & Mass of the gainer (in solar masses).\\
$M_\mathrm{d}\ [M_{\odot}]$ & $2.2\pm 0.2$                      & Mass of the donor (in solar masses).\\      
$R_\mathrm{g}\ [R_{\odot}]$  & $3.6\pm 0.2$                     & Mean radius of the gainer (in solar radii).\\
$R_\mathrm{d}\ [R_{\odot}]$  & $8.4\pm 0.2$                     & Mean radius of the donor (in solar radii).\\
$\log g_\mathrm{g}$          & $4.2\pm 0.1$                     & Logarithm of the gainer's effective gravity. \\
$\log g_\mathrm{d}$          & $2.9\pm 0.1$                     & Logarithm of the donor's effective gravity.\\
$M^\mathrm{g}_\mathrm{bol}$  &$-3.1\pm 0.2$                     & Absolute bolometric magnitude of the gainer. \\
$M^\mathrm{d}_\mathrm{bol}$  &$-1.9\pm 0.1$                     & Absolute bolometric magnitude of the donor.\\
$a_{orb}\ [R_{\odot}]$    & $29.7\pm 0.3$                       & Orbital semi-major axis. \\
${R}_\mathrm{disc}\ [R_{\odot}]$ & $12.9\pm0.3$                 & Disc radius (in solar radii).\\
$d_\mathrm{e}\ [R_{\odot}]$      & $0.6\pm 0.1$                 & Disc thicknesses at the edge in solar units. \\
$d_\mathrm{c}\ [R_{\odot}]$     & $0.3\pm 0.1$                  & Disc thicknesses at the centre in solar units.\\
            \hline
         \end{tabular} 
\tablefoot{Fixed parameters: $q=M_\mathrm{d}/M_\mathrm{g}=0.31$ -mass ratio of the components, $T_\mathrm{d}=9400 \mathrm{K}$ -temperature of the less-massive donor,
$F_\mathrm{d}=1.0$ -filling factor for the critical Roche lobe of the donor, $f_{\mathrm{g,d}}=1.00$ -synchronous rotation coefficients of the system components,
${\mathrm{\beta_{g,d}}=0.25}$ -gravity-darkening coefficients of the components, $A_\mathrm{g,d}=1.0$ -albedo coefficients of the components.}
\end{table*}

\begin{figure}
\includegraphics[]{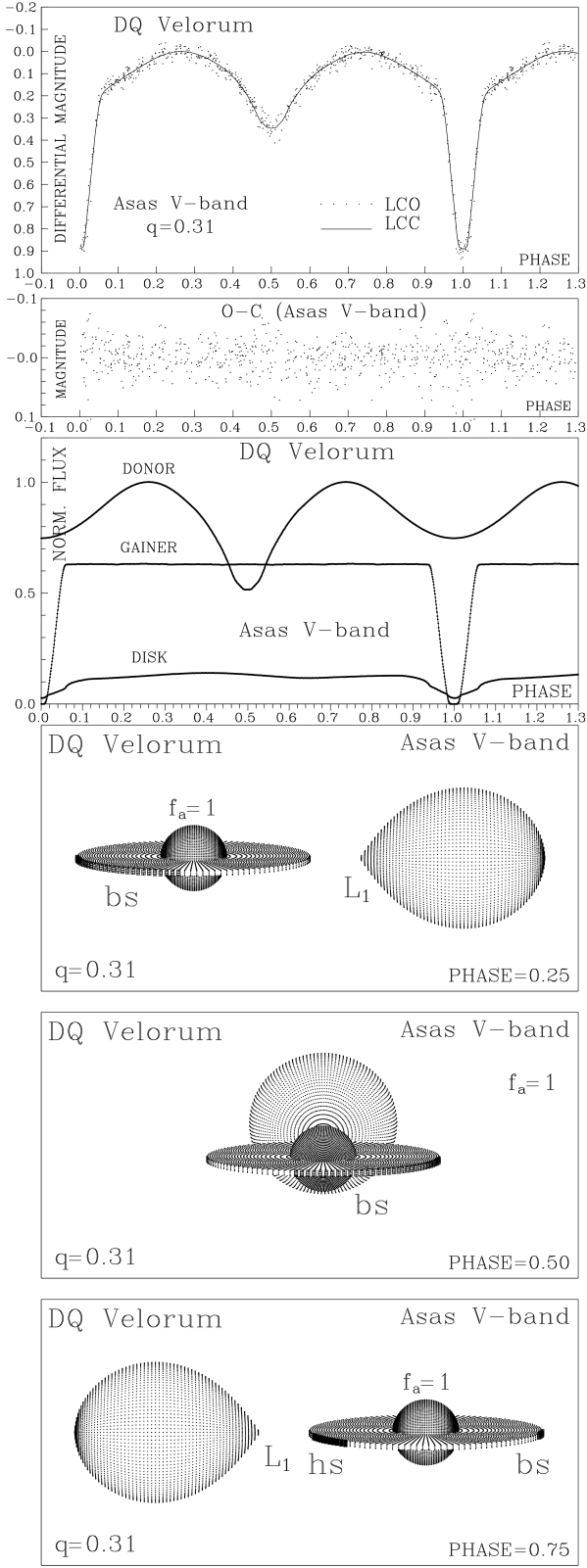}
\caption{Panels from top to bottom: Observed (LCO) and synthetic (LCC) light-curves of DQ Vel obtained by analysing photometric observations; final O-C residuals between the observed
and optimum synthetic light curves; fluxes of donor, gainer and of the accretion disc, normalised to the donor flux at phase 0.25; the views of the optimal model
at orbital phases 0.25, 0.50 and 0.75, obtained with parameters estimated by the light curve analysis.}
\end{figure}

\subsection{Possible pulsations of the B-type gainer star}
As suggested in Sect. 3.1, the scatter in the gainer RVs might be associated with pulsations of the B-type gainer star. The main reason to consider 
this scenario is the location of the gainer in the H-R diagram, as shown in Figure 12.
\begin{figure}
   \centering
   \includegraphics[width=6.4cm,angle=270]{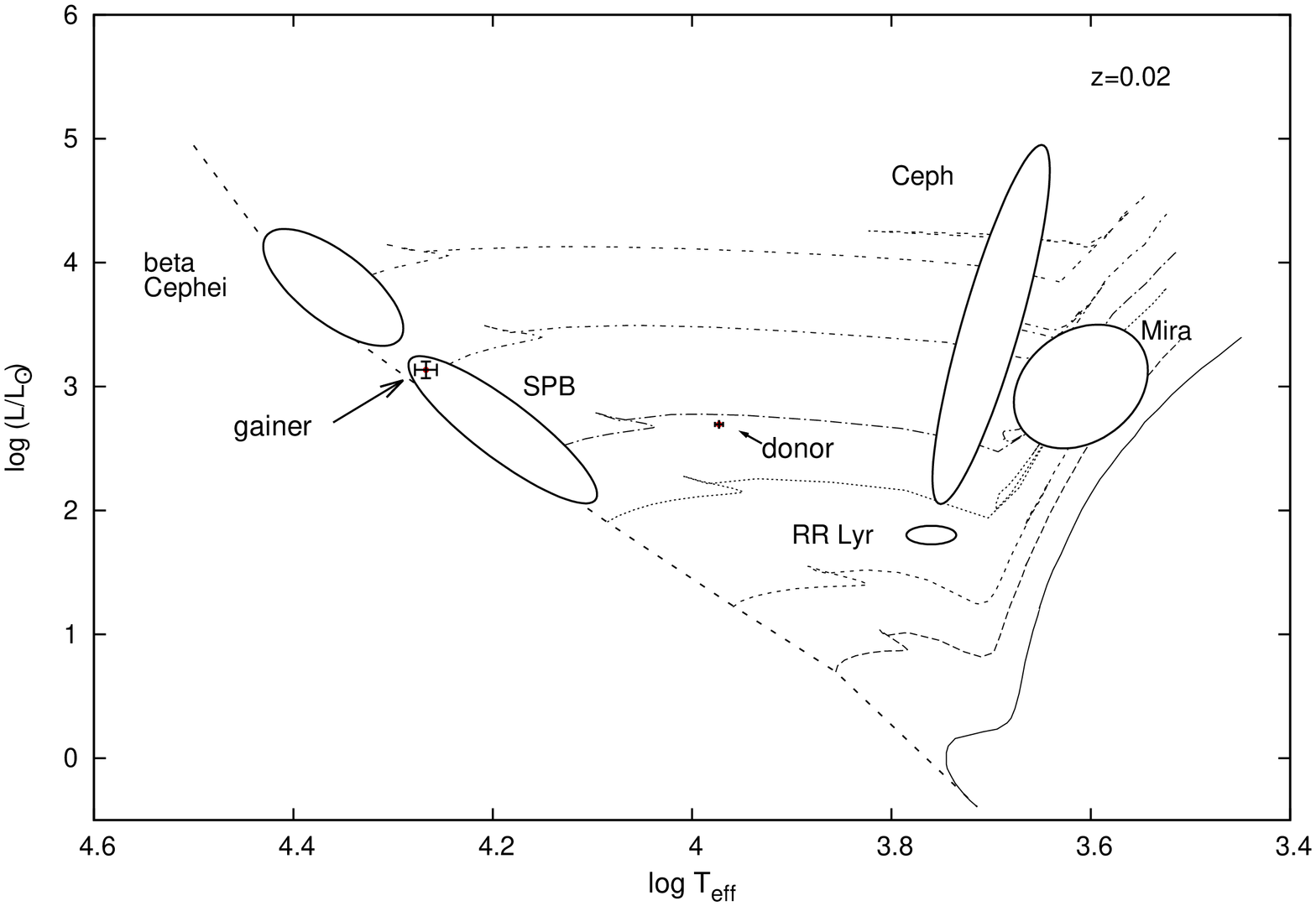}
      \caption{Location of DQ Vel including the errors of stellar components on the H-R diagram. Some stellar pulsation instability regions are indicated.
The evolutionary tracks were taken from Bressan et al. \cite{bressan} for z=0.02.}
   \end{figure}
According to its mass and luminosity, the gainer is located well within the instability region of slowly pulsating B-stars (SPBs).
These pulsators are typically B2 to B9 stars of intermediate mass between 3 to 7 $\mathrm{M_{\odot}}$. SPBs show multiperiodic light variations
of the order of days produced by high-order g-modes.
The typical photometric amplitude variations are lower than 0.1 magnitude while the radial velocity variations can reach up to 15 $\mathrm{km\,s^{-1}}$ (De Cat \cite{decata}, 
De Cat \& Aerts \cite{decatb}).\\
To find possible stellar pulsations, we carried out a Fourier frequency analysis on the gainer RVs and also on the O-C residuals of the 
V-band light curve after subtracting the orbital and long variabilities (see second panel in Figure 11).\\
The Fourier frequency analysis was performed with the software \textit{Period04} \footnote{http://www.univie.ac.at/tops/Period04/}, which is based on a discrete Fourier
 transform algorithm and is applied to the spectral 
analysis of unevenly sampled data.\\
We computed the Fourier transform (hereafter FT) for the gainer RVs data set. We found a significant peak of 0.16 $\mathrm{d^{-1}}$, which corresponds to the binary 
orbital frequency. In addition, several marginal peaks appear at higher frequencies, but the sampling of our scarce data set is not sufficient to 
unambiguously determine any periodicities above the noise level.\\
\begin{figure}
   \centering
   \includegraphics[width=9cm]{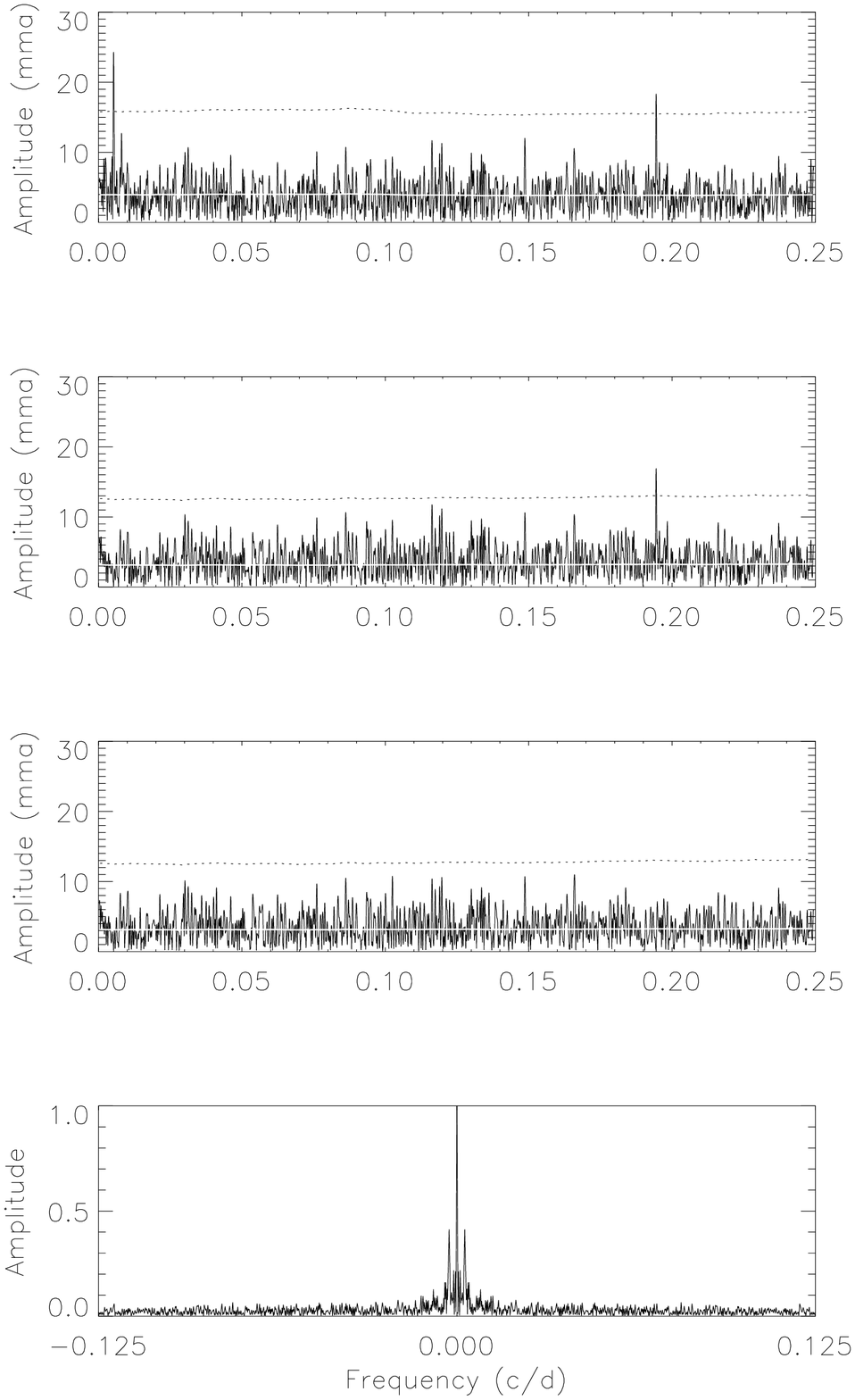}
      \caption{\textit{Top panel}: FT of the 516 data points obtained from the (O-C) DQ Vel photometric data. \textit{Middle upper panel}: FT of the residuals after 
prewhitening the original data by the first significant frequency ($f_{1}$). \textit{Middle lower panel}: FT of the residuals prewhitened by the two 
significant frequencies $f_{1}$ and $f_{2}$ listed in Table 7. The white solid line corresponds to $\sigma_{noise}$ and the dotted line corresponds to 4$\sigma_{noise}$.
\textit{Bottom panel}: Spectral window for the whole data set.}
   \end{figure}
Using the larger sample of 516 data points obtained after subtracting the orbital and long variability from the V-band light curve, we calculated the 
FT for this data set. In the top and bottom panels of Figure 13 we show the FT for the whole sample and its corresponding spectral window up to the Nyquist 
frequency ($f_{N}\approx0.25\,\mathrm{d^{-1}}$). To distinguish the significance of a frequency peak we used a 4$\sigma_{noise}$ criterion, which resulted in 
a 99.9\% confidence limit (Kuschnig et al. \cite{kuschnig}). To determine the $\sigma_{noise}$ in the Fourier spectrum we calculated the noise level
by averaging the amplitudes around each frequency along the whole frequency domain. As shown in Figure 13, the solid white line in the FT corresponds to this
mean noise level. The dotted line indicates a significance threshold of four times the $\sigma_{noise}$.  
After determining the highest frequency peak above 4$\sigma_{noise}$ in the FT of the original data set, we prewhitened the data, removing the sinusoid
corresponding to this frequency. Then, we calculated the FT of the residuals and found the next highest peak. We repeated this procedure
until no new peaks with significance above 4$\sigma_{noise}$ could be detected. 
In Table 7 we list the seven detected frequency signals. According to their S/N, only two frequencies ($f_{1}$ and $f_{2}$) 
are considered significant. In the middle upper panel of Figure 13 we show the FT of the residuals after removing $f_{1}$. Their corresponding $\sigma_{noise}$ 
and 4$\sigma_{noise}$ levels are shown. In the middle lower panel the prewhitened FT after removing $f_{1}$ and $f_{2}$ is displayed. No new 
significant peaks appear in the power spectrum after removing $f_{1}$ and $f_{2}$.
  \begin{table}
     \caption{Significant and marginal frequencies detected in the Fourier spectrum of the (O-C) DQ Vel photometric data.}
\centering
        \begin{tabular}{l c c c r}
            \hline
            ID  & frequency & amplitude & phase & S/N \\
                  & ($\mathrm{d^{-1}}$) & (mma) &  &   \\
            \hline
             $f_{1}$ & 0.005 & 24 & 0.65 & 6.04 \\
             $f_{2}$ & 0.194 & 17 & 0.43 & 5.18 \\
             $f_{3}^{*}$ & 0.166 & 11 & 0.82 & 3.67 \\
             $f_{4}^{*}$ & 0.086 & 11 & 0.38 & 3.67 \\
             $f_{5}^{*}$ & 0.030 & 11 & 0.37 & 3.56 \\
             $f_{6}^{*}$ & 0.148 &  9 & 0.14 & 3.08 \\
             $f_{7}^{*}$ & 0.102 &  9 & 0.74 & 2.95 \\
\hline
          \end{tabular}
\tablefoot{$^{*}$According to their S/N, these frequencies are considered marginal detections.}
   \end{table}
The frequencies $f_{1}=0.005\,\mathrm{d^{-1}}$ and $f_{3}=0.166\,\mathrm{d^{-1}}$ correspond to the long and 
orbital photometric variabilities, which are still present in the data. The origin of the significant frequency $f_{2}=0.194\,\mathrm{d^{-1}}$ 
corresponding to a period of 5.14 $d$ is unknown and could be interpreted as a pulsation of the SPB type.\\
We independently computed a simple Monte Carlo simulation to check the significance of $f_{2}$. To do this, we generated 1000 samples of synthetic random data with
a Gaussian distribution. For each sample, we generated an amount of points equal to the original number of 516 O-C data points and
used the same time sampling.
Then, we calculated the FT for each data set in the same frequency range as in the previous analysis, and found the amplitude of the highest frequency peak. 
In 99.9\% of cases we do not find any amplitude peak higher than $17\,\mathrm{mma}$. This is an additional independent test indicating the significance
of the $f_{2}$ detection.

\subsection{Distance to DQ Vel}
 It is well known that eclipsing double-lined binaries can be used as standard candles to calculate accurate distances for Local Group galaxies 
(Kang et al. \cite{kang2007}, Clausen \cite{clausen}). The most often used method for Milky Way binaries allows one to derive the distance modulus for either component
of a binary (Clausen \cite{clausen}). In the V-band we have 
\begin{eqnarray}
  (m_\mathrm{d,g}-M_\mathrm{d,g})_{0}  &=& 5\log(R_\mathrm{d,g}/R_{\odot})+(m_\mathrm{d,g}-A_\mathrm{v}) \nonumber \\
                         &-& M_\mathrm{bol\odot}+10\log(T_\mathrm{d,g}/T_{\odot})+BC_\mathrm{d,g} 
\end{eqnarray}
The absolute radii of stars and effective temperatures together with the visual apparent magnitudes of the components ($m_\mathrm{d},m_\mathrm{g}$), 
interstellar absorption ($A_\mathrm{v}$), and bolometric corrections ($BC$) are needed. Considering the apparent magnitudes on the light curves 
including the disc contribution, we determined the apparent magnitudes for the donor and gainer star using the V-band light curve and the 
theoretical fluxes from the photometric model (Fig. 11). 
From this last we measured the total flux of the system ($f_\mathrm{t}$) and the individual contributions ($f_\mathrm{d},\, f_\mathrm{g},\, f_\mathrm{disc}$),  
where $f_\mathrm{t}=f_\mathrm{d}+f_\mathrm{g}+f_\mathrm{disc}$
in the cuadrature phases $\phi_{o}=0.25$ and also $\phi_{o}=0.75$. The total flux at these orbital phases is related to the highest observed apparent magnitude 
of the system in the light curve $m_\mathrm{t}(V)=10.977$. 
Thus, the apparent magnitude of the donor $m_\mathrm{d}$ can be obtained from
\begin{eqnarray}
m_\mathrm{d}-m_\mathrm{t}&=& -2.5 \log \left(\frac{f_\mathrm{d}}{f_\mathrm{t}}\right) \nonumber \\
      m_\mathrm{d}&=& 11.578\pm0.01\,.
\end{eqnarray}
The same procedure was applied to the gainer star to obtain $ m_\mathrm{g}= 12.097\pm0.01$.
The bolometric corrections were obtained from the calibrations by Flower (\cite{flower}) using $\log T_\mathrm{g}=4.267$; $BC_\mathrm{g}=(-1.76\pm0.16)$ for the gainer 
and $\log T_\mathrm{d}=3.973$ and $BC_\mathrm{d}=(-0.13\pm0.05)$ for the donor. We determined the interstellar absorption $A_\mathrm{v}=1.02\pm0.02$ using the algorithm from
Cardelli et al. (\cite{cardelli}) that is available online \footnote{ $http://ned.ipac.caltech.edu/$}.\\ Equation 15 yields
\begin{equation}
 (m_\mathrm{g}-M_\mathrm{g})_{0}=12.46\pm0.23
\end{equation}

\begin{equation}
(m_\mathrm{d}-M_\mathrm{d})_{0}=12.47\pm0.09\,.
\end{equation}
Finally, we estimated the average distance from donor and gainer distance modulus:
\begin{eqnarray}
distance(pc)&= &10^{((m-M)_{0}+5)/5} \nonumber \\
distance(pc)&= & 3112\pm175 \,. \nonumber\\ 
\end{eqnarray}

\subsection{Comparison with V393 Scorpii}

During recent years a comprehensive study of the DPV system V393 Sco has been carried out by Mennickent at al. (\cite{mennickent2010b}, 
\cite{mennickent2012a}, \cite{mennickent2012b}) to study the nature and evolution of the system. The studies include photometry 
(VIJK bandpasses and V-ASAS light curves) and high-resolution spectroscopy in the optical, near-infrared, and ultraviolet wavelength ranges. 
V393 Sco is a galactic ($d=523\ \mathrm{pc}$) bright DPV composed of an A-type donor ($T_\mathrm{d}= 8600\pm600\ \mathrm{K}$) and a B-type gainer ($T_\mathrm{g}= 
16600\pm500\ \mathrm{K}$) 
surrounded by a massive disc. The system shows a complex variability of spectral features along the orbital and long cycles including intense 
double-peaked emission in optical Balmer lines.
\begin{table}
     \caption{Comparison of the main stellar and disc parameters for DQ Vel and V393 Sco.
 }
\centering
        \begin{tabular}{l l r}
            \hline
            Parameter  & DQ Vel & V393 Sco\\
            \hline
              & \emph{Stellar Components} & \\
             \hline
             $T_\mathrm{g} [\mathrm{K}]$  &  $18\,500\pm500$ &  $16\,600\pm500$ \\
             $T_\mathrm{d} [\mathrm{K}]$  &  $9400\pm100$ &  $8600\pm600$      \\
             i $\mathrm{[^{\circ}]}$   &  $82.5\pm0.2$ & $80.0 \pm 0.2$   \\
             $M_\mathrm{g}\ [M_{\odot}]$ & $7.3 \pm 0.3$ &  $7.8 \pm 0.2$ \\
             $M_\mathrm{d}\ [M_{\odot}]$  & $2.2 \pm 0.2$ &  $2.0 \pm 0.2$ \\
             $R_\mathrm{g}\ [R_{\odot}]$  & $3.6 \pm 0.2$ & $4.4 \pm 0.2$ \\
             $R_\mathrm{d}\ [R_{\odot}]$  & $8.4 \pm 0.2$ & $9.4 \pm 0.3$ \\
             $\log g_\mathrm{g}$ & $4.2\pm0.1$ & $4.0\pm0.1$ \\
             $\log g_\mathrm{d}$ & $2.9\pm0.1$ & $2.8 \pm 0.1$\\
             $a_\mathrm{orb}\ [R_{\odot}] $ & $29.7\pm0.3$ &  $35.1pm0.3$ \\
            \hline
               & \emph{Disc}& \\
            \hline
            $F_\mathrm{disc}$       &  $0.89 \pm 0.03$ & $0.55 \pm 0.04$     \\
            $R_\mathrm{disc}\ [R_{\odot}]$ & $12.9 \pm 0.3$ &  $9.7 \pm 0.3$   \\  
            $d_\mathrm{e}\  [R_{\odot}]$  & $0.6 \pm 0.1$ & $1.3 \pm 0.3$ \\
            $d_\mathrm{c}\  [R_{\odot}]$  & $0.3 \pm 0.1$ & $2.1 \pm 0.4$\\
            \hline
          \end{tabular}
\tablefoot{Results for V393 Sco were taken from Model A in Mennickent et al. (\cite{mennickent2012a}). Parameters for DQ Vel are taken from this work (Table 6).}
   \end{table}
\\
We noted strong similarities of the physical properties of V393 Sco and DQ Vel, as can be seen in Table 8. Physical parameters of the stellar components such as masses,
radii, temperatures, gravities, and also orbital parameters (binary separation and inclination) are similar. However, a detailed comparison showed significant
differences in the optical spectral features of the two systems. We suggest that these differences are associated with the physical properties of the accretion discs
and thus with the mass transfer rates. As shown in Table 8, the disc in DQ Vel seems to be more extended and colder than that of V393 Sco. Moreover, the hot and bright spots in V393 Sco are
hotter than in DQ Vel. From the geometric point of view, DQ Vel's disc is concave and thick at the outer border, in contrast to the modelled disc in V393 Sco, which 
is convex and probably has most of its mass concentrated close to the gainer. We suggest that the main differences of the discs are related to different evolutionary stages.
The high mass-transfer rate of V393 Sco would form a massive accretion disc and lead the gainer to rotate critically. DQ Vel instead could be an older 
system in a state of lower mass transfer where the gainer had time to slow down, allowing the formation of an extended disc.
A detailed analysis of the evolutionary stage of DQ Vel remains to be conducted in the future.

\section{Summary}
Our analysis of VIJK (REM-telescope) and V (ASAS) photometric data togheter with the study of 46 optical high-resolution echelle spectra along several orbital cycles allowed 
us to estimate the main physical properties for the components of the DPV system DQ Vel.\\
We separated the composite spectra using the Doppler shifts of known spectral lines to obtain two single-lined templates, which were used to derive the 
RV curves through a Fourier cross-correlation process, and obtained the spectroscopic mass ratio ($q=0.31\pm0.03$) of the system. 
A comparison of the templates with a grid of synthetic spectra enabled us to estimate the effective temperature $T_\mathrm{d}=9400\pm100\ \mathrm{K}$ for 
the donor star.
We calculated the fractional light contribution of the donor star to subtract it from the spectra and thus analyse the gainer spectral
features. The donor-subtracted spectra along the orbital cycle show a variability of the Balmer and helium profiles, suggesting the presence of circumstellar
material. The spectroscopic mass ratio and donor temperature were used to carry out a V-band light curve fitting and to obtain the main binary elements 
such as the masses $M_\mathrm{g}=7.3\pm0.3\ M_{\odot}$ and $M_\mathrm{d} =2.2\pm0.2\ M_{\odot}$, and the radii $R_\mathrm{g}=3.6\pm 0.2\ R_{\odot}$ and 
$R_\mathrm{d}=8.4\pm0.2\ R_{\odot}$ for gainer and donor star together with the orbital separation ($a_\mathrm{orb}=29.7\pm0.3\ R_{\odot}$) 
and the inclination angle ($i^{\mathrm{\circ}}=82.5\pm0.2$) of the system.
The best model suggest a semi-detached system comprising an A1III donor star ($T_\mathrm{d}=9400\pm100\  \mathrm{K}$) and a B3V gainer star ($T_\mathrm{g}=18\,500\pm500\ \mathrm{K}$)
with an extended accretion disc around the gainer. The model includes two active regions (hot and bright spots) located at the outer edge of the disc.
The temperatures and radii of the stellar components together with the individual apparent magnitudes determined from the light curve were used to estimate 
the distance to DQ Vel ($d=3.112\ \mathrm{kpc}$).\\
Based on the position of the gainer on the HR diagram, which is well within the SPB instability strip, we interprete this periodicity of
$5.14\,d$ as a pulsation of an SPB type formed in the gainer.\\
Similarities on the stellar parameters for DQ Vel and the DPV system V393 Sco were found during this study. However, we observed 
differences on the geometrical and physical properties of the two accretion discs, and we suggest that they are related to different evolutionary stages. 
We are currently working on a detailed analysis of the evolutionary stage of DQ Vel and aim to present our results in an forthcoming publication.
 
\begin{acknowledgements}
R. M. acknowledges support by Fondecyt grant 1110347 and from the BASAL Centro de Astrof\'isica y Tecnologias Afines (CATA) PFB--06/2007. 
G. D. gratefully acknowledges the financial support of the Ministry of Education and Science of the Republic of Serbia through the project 176004, 
\textquotedblleft Stellar physics \textquotedblright.
D. B. acknowledges support by the Chilean CONICYT PhD grant and the European Southern Observatory grant.

\end{acknowledgements}

\end{document}